%% file: main.tex
\documentclass[10pt,conference]{src/IEEEtran}
\input{src/packages.tex}

\input{src/macros.tex}

\begin{document}

\title{FaaSter Troubleshooting - Evaluating %
Distributed Tracing Approaches %
for Serverless Applications}
\author{\IEEEauthorblockN{Maria C. Borges}
\IEEEauthorblockA{\textit{Information Systems Engineering} \\
\textit{Technische Universität Berlin}\\
Berlin, Germany \\
mb@ise.tu-berlin.de}
\and
\IEEEauthorblockN{Sebastian Werner}
\IEEEauthorblockA{\textit{Information Systems Engineering} \\
\textit{Technische Universität Berlin}\\
Berlin, Germany \\
sw@ise.tu-berlin.de}
\and
\IEEEauthorblockN{Ahmet Kilic}
\IEEEauthorblockA{
\textit{Technische Universität Berlin}\\
Berlin, Germany \\
ak@ise.tu-berlin.de}
}

\maketitle
\begin{abstract}
   \input{sections/00_abstract.tex}

\end{abstract}

\begin{IEEEkeywords}
Serverless Computing, FaaS Platforms, Distributed Tracing, Observability
\end{IEEEkeywords}

\section{Introduction}\label{sec:intro}
\input{sections/01_introduction}

\section{Problem Statement}\label{sec:problem}
\input{sections/02_problem}

\section{Fault Observability}\label{sec:model}
\input{sections/03_sdom}

\section{Design and Implementing of Serverless Tracing Approaches}\label{sec:implementation}
\input{sections/04_implementation}

\section{Experimental Evaluation}\label{sec:eval}
\input{sections/05_eval}

\section{Related Work}\label{sec:relatedwork}
\input{sections/06_relatedwork}

\section{Conclusion}\label{sec:conclusion}
\input{sections/07_conclusion}

\bibliographystyle{src/IEEEtran}
\bibliography{src/IEEEabrv,references_cleaned_noURLs}

\end{document}

%% file: src/packages.tex
\usepackage{afterpage}
\usepackage{algorithmic}
\usepackage{amsmath,amssymb,amsfonts}
\usepackage{booktabs} %
\usepackage{cite}
\usepackage{color}
\usepackage{graphbox}
\usepackage{graphicx}
\usepackage[breaklinks=true]{hyperref}

\usepackage{mathabx}
\usepackage{mathtools}
\usepackage{multirow}
\usepackage{url}
\usepackage{tabularx}
\usepackage{textcomp}
\usepackage[table,xcdraw,dvipsnames]{xcolor}
\usepackage{hhline}
\usepackage{colortbl}
\usepackage[normalem]{ulem}
\usepackage{cleveref} %

\usepackage{listings}

\usepackage{flushend}
\usepackage{diagbox}

%% file: src/macros.tex
\newcommand{\cbstart}{}
\newcommand{\cbend}{}

\makeatletter
\newcommand{\hourglass}{}%
\DeclareRobustCommand{\hourglass}{\mathrel{\mathpalette\hour@glass\relax}}

\newcommand\hour@glass[2]{%
  \vcenter{\hbox{%
    \rotatebox[origin=c]{90}{$\m@th#1\bowtie$}%
  }}%
}
\makeatother

%% file: sections/00_abstract.tex
Serverless applications can be particularly difficult to troubleshoot, as these applications are often composed of various managed and partly managed services. 
Faults are often unpredictable and can occur at multiple points, even in simple compositions.
Each additional function or service in a serverless composition introduces a new possible fault source and a new layer to obfuscate faults. 
Currently, serverless platforms offer only limited support for identifying runtime faults. 
Developers looking to observe their serverless compositions often have to rely on scattered logs and ambiguous error messages to pinpoint root causes. 

In this paper, we investigate the use of distributed tracing for improving the observability of faults in serverless applications. 
To this end, we first introduce a model for characterizing fault observability, then provide a prototypical tracing implementation—specifically, a developer-driven and a platform-supported tracing approach.
We compare both approaches with our model, measure associated trade-offs (execution latency, resource utilization), and contribute new insights for troubleshooting serverless compositions.

%% file: sections/01_introduction.tex
Building cloud-based applications involves selecting and combining diverse technologies, various service offerings, and a multitude of development frameworks. Given these circumstances, observing applications for debugging and fault detection purposes can quickly become a challenging task. 

The lack of observability in cloud-based applications is especially apparent in fully managed services like serverless platforms~\cite{paper_werner_et_al_Diminuendo}. These platforms enable workload driven, elastic utilization of resources, with FaaS being the most prominent example. In FaaS, developers build function compositions by uploading function code and specifying triggering events. Functions are then provisioned and run on-demand, avoiding the need for developers to manage their own servers. 

However, by relinquishing control of the infrastructure, developers also lose control over the execution and have to depend on the limited options provided by these platforms to observe and inspect an application.
According to Manner et al.~\cite{2019-Manner-SICS-FaasDebugging}, Hellerstein et al.~\cite{2019-Hellerstein-CoRR-ServerlessComputing}, and Kuhlenkamp et al.~\cite{kuhlenkamp-2020-ic2e-ifs_and_buts}, this loss of control remains one of the main challenges in serverless application development. 
Thus, while serverless computing platforms like AWS Lambda or OpenWhisk offer many benefits for the developer, these platforms only offer limited support to identify and troubleshoot runtime faults. 

In this paper, we argue that it is important for application developers to consider the observability of their serverless applications. 
Today, observability is typically achieved through response messages and logging services, which we show is not enough. 
Consequently, we investigate the use of distributed tracing for improving the observability of faults in serverless applications. 
With our work, we provide the first contributions to answer the research question: \textbf{How can application developers characterize the qualities and associated cost of fault observability systems for serverless applications?}

Towards this end, we present the following contributions:
\begin{itemize}
    \item %
    A serverless fault observability model and a first instantiation of the model on the example of AWS Lambda and OpenWhisk.
    \item %
    A prototypical implementation and preliminary evaluation of two serverless tracing approaches to improve fault observability in OpenWhisk. %
\end{itemize}

The remainder of this paper is structured as follows:
First, section~\ref{sec:problem} refines our initial problem statement. We propose a fault observability model in section~\ref{sec:model} and apply it to common faults in serverless compositions. We present the design and implementation of two observability approaches in section~\ref{sec:implementation}. In section~\ref{sec:eval}, we evaluate both implemented approaches and the applicability of tracing in serverless applications in general. Lastly, we present related work in section~\ref{sec:relatedwork} and the conclusion in section~\ref{sec:conclusion}.

%% file: sections/02_problem.tex
Pinpointing the cause of a failure in a distributed system can be a notoriously difficult task. Due to the inherent distribution, it is not always clear where a failure originates, plus potential problems are varied and unpredictable~\cite{2015_MaceFonseca_SOSC_PivotTracing}. Serverless applications are no different, as they are essentially compositions of multiple independent distributed systems. 
Even the most basic setup with one function implies a composition of FaaS-function and API-Gateway, and a call between these two services.
Most applications typically go beyond a single function and include calls to other functions or to external services, e.g. object storage services. 

\begin{figure}[b]
    \centering
    \includegraphics[width=\columnwidth]{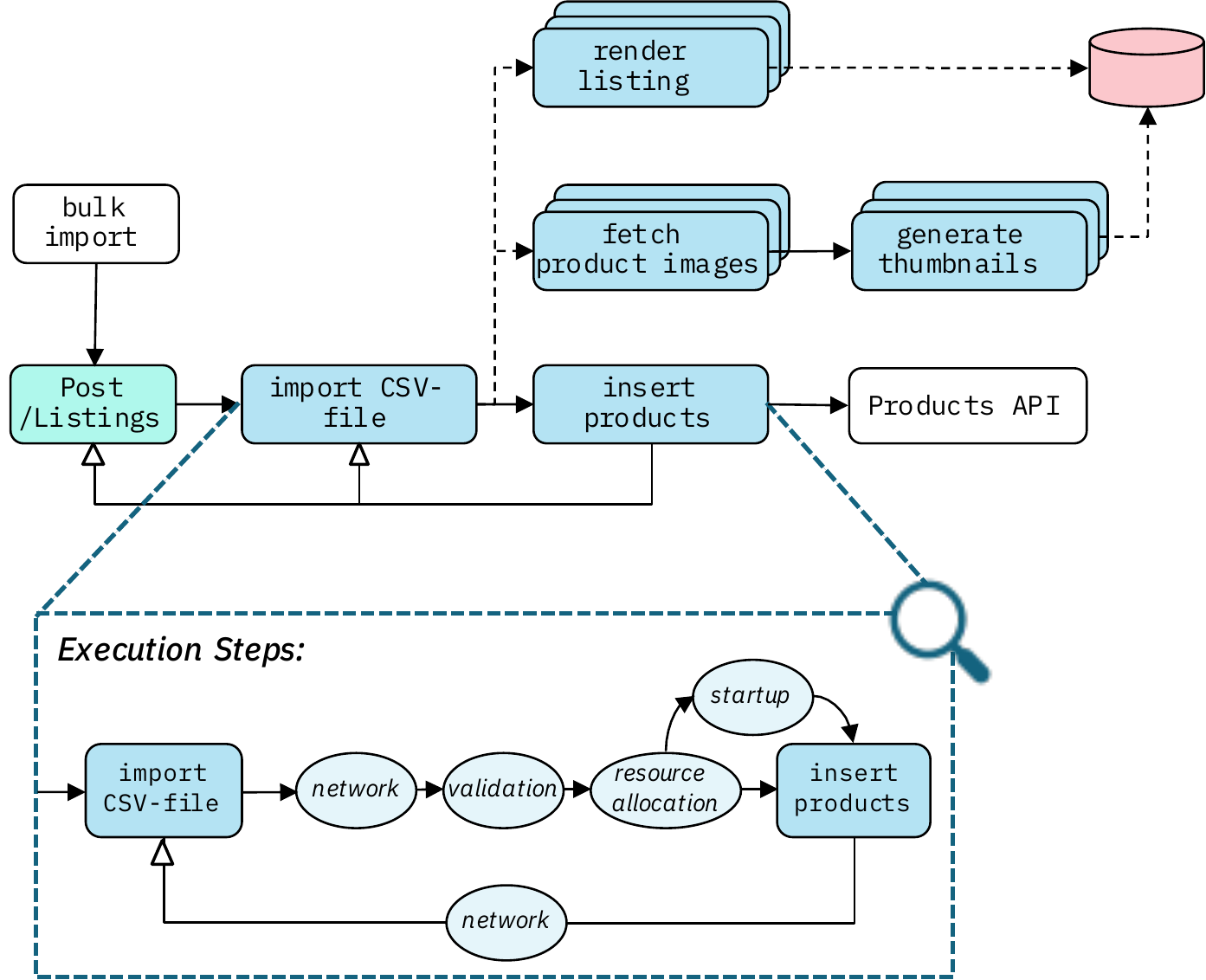}
    \caption{Example serverless composition  of a  bulk import feature in a e-commerce marketplace application with a detailed view of the different execution steps in the execution path.}
    \label{fig:serverless-example}
    \label{fig:app}
\end{figure}

Figure \ref{fig:serverless-example} illustrates a segment of an exemplary serverless e-commerce marketplace application. In this serverless composition, the client calls the shop backend to add new product listings to the store. The API Gateway routes the request to the appropriate import function. Once this function is finished executing, it triggers other functions. One synchronous function is executed to update the product catalogue and corresponding API. Additionally, each product listing also spawns asynchronous function executions that handle thumbnails and listing rendering. If nothing goes wrong, a call to the serverless application will traverse every component in the composition, persist the data in the storage services, and return a successful response.

However, faults are unavoidable in distributed systems, so failures will inevitably occur. As Werner Vogels puts it, "everything fails all the time"~\cite{2020_AmazonCTOInterview_EverythingFails}.
Here, we consider a \textit{failure} any type of execution that visibly deviates from the specified behaviour~\cite{2004_FaultLocalizationSurvey_old}, i.e. a timeout or an error. 
A failure is the consequence of a \textit{fault}. 
Sometimes faults are introduced by the development team, for example if they deploy a faulty function.
Other times, they are beyond the developer's control and lie with the serverless platform, for example if a network error occurs.
In any case, it is important that the development team isolates the fault or cause of a failure to be able mitigate it and ultimate deliver a reliable application. 
Developers rely on \textit{evidence} to recognize failures and isolate faults. 
Some faults are easily observable because the evidence is unambiguous. 
Other faults are only partially observable, when the evidence is not enough to precisely locate the fault, or when the evidence is obfuscated because the failure appeared far away from the source of the fault.

Each additional component in a serverless execution introduces a new source for faults. Even simple compositions can produce inconsistent evidence in the event of a fault. For example, the log of our \texttt{insert products} function might show an error message due to a developer error, while the upstream \texttt{import CSV-file} function sees a timeout instead. 
Moreover, faults can occur not only during the execution of the function, but also in the preceding execution steps, like validation, resource allocation, or startup of a new runtime for the function (see also figure~\ref{fig:serverless-example}). This again can lead to fault ambiguity, for example a timeout can be the result of a network error but also the result of delayed startup, referred to as a cold-start~\cite{manner_cold_2018}. 

While the development team can use knowledge of the application architecture to identify possible
culprits, determining the exact cause usually requires understanding how the application behaves at runtime. 
Observability systems like response message monitoring or logging services expose runtime behaviour and generate evidence, so they are frequently used for troubleshooting serverless applications. 
Beyond these simple approaches, distributed tracing is increasingly being considered as an alternative for serverless fault observability~\cite{Lin_TracingLambdaDependencies_2018}. Tracing enables end-to-end observation of application behaviour by retrieving and aggregating log data in so-called \textit{traces}~\cite{2010_Sigelman_Dapper}.

Developers need troubleshooting tools that can help them identify faults in serverless applications. While the offering of observability systems continues to increase, it is hard to know what advantages and associated costs different observability approaches entail. We argue that a clear understanding of faults and fault observability is required, both for developers looking to observe their serverless applications and for platform providers looking to implement new observability tooling. Thus, we formulate the following two research problems:
\begin{itemize}
    \item \textbf{P1: How can we classify fault observability in serverless applications and serverless platforms?}
    \item \textbf{P2: How can we implement and evaluate fault observability tooling in serverless systems?}
\end{itemize}

%% file: sections/03_sdom.tex
In this section, we present our fault observability model.
We introduce the concept of fault evidence as the basis for our model, collect common fault scenarios in serverless applications (without claim of completeness) and apply the model to these examples.

\subsection{Fault Observability Model}

Our model characterizes the observability of a fault by formalizing the factors that accelerate fault identification.
Specifically, we focus on fault information, i.e. any data produced by the system related to the fault occurrence, which we label as \textit{evidence}. 
Fundamentally, the model takes into account three factors for fault observability: (1) \textit{evidence visibility}, (2) \textit{evidence ambiguity}, and (3) \textit{evidence inconsistency}.

\cbstart
\textit{Evidence visibility} denotes the fact that evidence of a fault may become visible in different points during an application life-cycle. 
Notably, the evidence of a fault may be visible in the response to a request at runtime, visible in a log file after execution or only visible when the execution path of a request is analysed through tracing. 
Thus, we characterize evidence visibility as \textit{response-observable}, \textit{log-observable} and \textit{trace-observable}. 
\textit{Response-observable} describes evidence that can be observed in the response of a function invocation. This includes both the response code and the response body. 
\textit{Log-observable} describes evidence that can be found in the logs of the different functions, and  \textit{trace-observable} describes evidence that can be observed through request tracing. 
Depending on the visibility, detection and mitigation strategies become more complex and time-consuming for a developer.
\cbend

\cbstart
\textit{Evidence ambiguity} stems from the fact that the same evidence may be generated as an indication of more than one fault~\cite{2004_FaultLocalizationSurvey_old}. 
Evidence can therefore be \textit{ambiguous} or \textit{unambiguous} within its visibility. 
Intuitively, faults that produce \textit{unambiguous} and \textit{response-observable} evidence are easier and faster to troubleshoot, e.g. responses with an error message that point to the exact fault.
Faults that are \textit{ambiguous} in their response-observable evidence require developers to look at other evidence channels, such as log-observable evidence or trace-observable evidence. 
An example here are faults that result in a timeout. 
A timeout is response-observable evidence, because the API Gateway delivers a response denoting a function-execution timeout. 
However, it is ambiguous evidence, because the developer still has to rule out many possible candidates in the composition to find the fault. 
Through log inspection, developers are able to reduce ambiguity, i.e. reduce the number of fault candidates, but evidence can still be ambiguous within log-observable visibility, i.e. when the same log entries are generated for different faults. 
Further, correlation of log evidence across multiple function log files is typically only possible when faults don't result in a timeout because custom error handling code can't be executed then, so timeouts in particular tend to produce ambiguous evidence. 
\cbend

\textit{Evidence inconsistency} occurs when the event of a fault is misrepresented or obfuscated by part of the system. Sometimes this is caused by improper reporting of a visible and observed fault. Other times, it results from a disagreement between different components, due to the decoupled nature of the execution. 
When a fault occurs further down in the execution path, the evidence generated by downstream functions is not always propagated upstream, so upstream functions may report the system as operating correctly.
Evidence is considered \textit{consistent} when a fault is accurately reported and not obscured by upstream services.
Inconsistencies can delay fault detection by hiding  and lead developers on the wrong path during fault localization.
Asynchronous calls to downstream functions are especially prone to inconsistencies. Here, developers might be misled by a successful response message after the execution of the synchronous dependencies, even though the asynchronous function failed downstream.

With this foundation, we extrapolate the following definition for fault observability of a system:
The observability of a system is dependent on its ability to produce evidence that is \textit{visible, consistent, and unambiguous}. The better a system can enable these properties, the more observable faults in the system become.

In the remainder of this section, we apply this model to serverless application faults and discuss the observability of faults in AWS Lambda and OpenWhisk without distributed tracing. %

\begin{table}[tp]
    \caption{Examples of faults that can occur in function compositions}
    \input{data/faults_v4}
    \label{tab:faults}
\end{table}

\begin{table}[b]
    \caption{Response-observable evidence in AWS Lambda and OpenWhisk for fault scenarios in \ref{tab:faults}}
    \input{data/response_observability}
    \label{tab:response-observability}
\end{table}

\begin{table}[tp]
    \caption{Log-observable evidence in AWS Lambda and OpenWhisk for faults scenarios in \ref{tab:faults}}
    \input{data/log_observability}
    \label{tab:log-observability}
\end{table}

\subsection{Applying the Model: Serverless Fault Observability}\label{sec:appliedmodel}
\noindent Platforms can vary in their observability of serverless applications, as each platform handles faults differently and offers different tooling. While AWS Lambda might attempt to retry functions when an unforeseen error occurs, other platforms like OpenWhisk fail and report an error instead. In both cases, it is crucial to use platform knowledge to understand the ambiguity and inconsistency of evidence when debugging serverless applications.

In Table~\ref{tab:faults} we present four typical fault scenarios in serverless applications. %
We apply our fault observability model to these scenarios in Table~\ref{tab:response-observability} and \ref{tab:log-observability}, relying on experiments and documentation to fill out the table.
Our model shows how faults differ in evidence visibility, and how different platforms behave in regards to evidence ambiguity and inconsistency. The model also serves as the basis for the evaluation in section~\ref{sec:eval}, where we complete the table with trace-observable evidence. 

First, in Table \ref{tab:response-observability}, we look at response-observable evidence, i.e. any information included in the response of a function invocation. Here, we distinguish between the response code (aka StatusCode) and the response body. Evidence is considered inconsistent when the response body doesn't match the response code. 
This is the case for \texttt{F1-F3} in AWS, as the platform always returns a successful 200 response code by default. This inconsistency can have several implications for applications developers. In contrast, OpenWhisk produces consistent response evidence for the first three scenarios (5xx status code and error message).
Apart from revealing this AWS quirk, the table also highlights the value and shortcomings of response-observable evidence. Response messages can be an adequate troubleshooting mechanism for faults like \texttt{F1}, that are self-contained and explicit in their error messages. 

However,  other faults are not accurately observed by this system, for example faults that result from the complex interplay between components .
For \texttt{F2} and \texttt{F3}, only a timeout error response is returned, which is ambiguous evidence.
\texttt{F4} is another case where response-observable evidence is insufficient and inconsistent.  The fault in the asynchronously called function will not be registered by the function upstream, who returns a successful response regardless.

Using the results from Table~\ref{tab:response-observability} as the baseline, we can now evaluate what observability improvement log services provide over basic response messages.
We apply our model to log-observable evidence in Table~\ref{tab:log-observability}. In contrast to response-observable evidence, which relies on a single response to expose the behaviour of the serverless application, log-observable evidence can be sourced from multiple components in the serverless composition.
The fault scenarios described in Table~\ref{tab:faults} only include compositions with max. two components, so evidence is either recorded in an upstream log or a downstream log. Improvements over response-observable evidence are underlined in the table. 

Unsurprisingly, our results show improvements for evidence consistency in AWS. Log evidence does not suffer from the same quirk as the response code, so evidence is now consistent for \texttt{F1} and \texttt{F2}. 
Another improvement is the unambiguous fault evidence collected in the log file of the asynchronous function \texttt{F4}. However, \texttt{F4} is still obfuscated by the upstream function, who doesn't notice the fault and continues to log successes. Evidently, log services don't appear to provide many benefits over response messages, and are unsuitable for troubleshooting scenarios \texttt{F2-F4}.

In summary, we can see in these examples that pin-pointing faults using only log-observable and response-observable evidence can present challenges. Executions that fail with a timeout produce ambiguous response-observable and log-observable evidence, which is hard to troubleshoot due to the multitude of possible causes. Errors in asynchronous function can easily go unnoticed and generate inconsistent evidence.
Thus, developers require tools that improve both consistency and reduce the ambiguity of observable evidence. %
Tracing addresses these issues by exposing the behaviour of functions and services across the entire execution path. 
However, unlike logging, tracing is not always supported by platform providers (e.g. OpenWhisk), or not enabled by default (AWS offers the ability to trace Lambda after enabling X-Ray). 
Thus, developers must either use workarounds to implement distributed tracing in their business logic, or hope that the tracing solution offered by the platform provider fulfils their requirements.

In the remainder of the paper, we evaluate the benefits and trade-offs associated with tracing serverless applications. Using the model first demonstrated in Table~\ref{tab:response-observability}, we  analyse the trace-observable evidence produced by two different approaches and see how they fare with regard to evidence ambiguity and evidence inconsistency. Lastly, we also measure associated performance and cost trade-offs for both developers and platform providers.

%% file: data/faults_v4.tex
\centering
\begin{tabularx}{\columnwidth}{|l|X|}
        \hline
        Fault   &   Description \\   
        \hline
        F1    &   The function is called but fails due to a developer bug  \\
        F2    &   The function calls a third-party API and times out before receiving a response   \\
        F3      &   The upstream function times out after a synchronous call to a cold downstream function \\
        F4      &   After the upstream function concludes, an asynchronously triggered function fails downstream due to a developer bug  \\
        \hline
\end{tabularx}

%% file: data/response_observability.tex
\begin{tabular}{|ll|ll|cc|} 
\hline
\multirow{2}{*}{Platform} & \multirow{2}{*}{} & \multicolumn{2}{c|}{evidence} & \multicolumn{1}{c}{\multirow{2}{*}{consistent}} & \multirow{2}{*}{unambiguous}  \\
                          &                   & resp-code & resp-body         & \multicolumn{1}{c}{}                            &                               \\ 
\hhline{|======|}
\multirow{4}{*}{AWS}      & F1                & success   & error             & false                                           & true                          \\
                          & F2                & success   & error (TO)        & false                                           & false                         \\
                          & F3                & success   & error (TO)        & false                                           & false                         \\
                          & F4                & success   & success           & false                                           & -                             \\ 
\hline
\multirow{4}{*}{OWhisk}   & F1                & error     & error~            & true                                            & true                          \\
                          & F2                & error     & error (TO)        & true                                            & false                         \\
                          & F3                & error     & error (TO)        & true                                            & false                         \\
                          & F4                & success   & success           & false                                           & -                              \\
\hline
\end{tabular}

%% file: data/log_observability.tex
\begin{tabular}{|ll|ll|cc|} 
\hline
\multirow{2}{*}{Platform} & \multirow{2}{*}{} & \multicolumn{2}{c|}{evidence} & \multirow{2}{*}{consistent} & \multirow{2}{*}{unambiguous}  \\
                          &                   & upstream   & downstream       &                             &                               \\ 
\hhline{|======|}
\multirow{4}{*}{AWS}      & F1                & error      & n.a.             & \uline{true}                        & true                          \\
                          & F2                & error (TO) & n.a.             & \uline{true}                        & false                         \\
                          & F3                & error (TO) & success          & false                       & false                         \\
                          & F4                & success    & error            & false                       & \uline{true}                          \\ 
\hline
\multirow{4}{*}{OWhisk}   & F1                & error      & n.a.             & true                        & true                          \\
                          & F2                & error (TO) & n.a.             & true                        & false                         \\
                          & F3                & error (TO) & success          & false                       & false                         \\
                          & F4                & success    & error            & false                       & \uline{true}                          \\
\hline
\end{tabular}

%% file: sections/04_implementation.tex
Based on the discussion presented in the previous sections, we see that the correlation of fault events is necessary to address evidence ambiguity and inconsistencies in troubleshooting serverless applications.
Distributed tracing systems like Dapper~\cite{2010_Sigelman_Dapper} and other subsequent industry implementations enable end-to-end observation of microservice applications by retrieving and aggregating log data in so-called \textit{traces}. 
Here, a single trace consists of a set of causally related events, which demarcate meaningful points of interaction in the execution path of a request. 
Causality between events is established with a correlation identifier (\textit{traceID}), which is propagated through the system, implying that all parts of the system must interact with the tracing system through tracing instrumentation.

Thus, we explored two ways to implement and instrument serverless applications with these tracing approaches in OpenWhisk\footnote{https://github.com/apache/openwhisk}, one through developer-driven tracing and one with platform-supported tracing. 
Both approaches assume that a developer can instrument the business logic of the application. 
However, these approaches differ in the responsibilities of providing the tracing infrastructure and the level of detail available in traces.

Further, tracing can quickly lead to large volumes of data, especially if every single request in the application is traced, which can be expensive to collect, store and process. This can potentially impact both application and platform quality. We therefore consider the impact to both application and platform when evaluating these two design options.

\subsection{Developer-driven Tracing}
Developer-driven tracing in a serverless application is implemented as part of the function code. 
Multiple libraries and tools exist that support some of the existing FaaS platforms\footnote{https://thrunder.io}.
We enriched each implementation with instrumentation commands throughout the execution path. 
Before a function can return, it must write out the tracing data to an external backend. Thus, these approaches require deploying and running a secondary backend to capture the spans and require some overhead in function executions to persist trace information.

We implemented a custom Python module for demonstration and evaluation purposes to instrument OpenWhisk functions for the Zipkin backend.%
We opted to implement our own module because none of the existing frameworks is flexible enough to let us configure the Zipkin collector endpoints.
Furthermore, we observed that most existing frameworks, e.g., \texttt{cppkin, py\_zipkin, and flask\_zipkin} only wrap the Zipkin API.
Therefore, we followed the same approach and sent spans to the Zipkin API via Rest/HTTP. In addition, the same module can be reused for our implementation of platform-supported tracing. 

One challenge of developer-driven tracing is that the function code is also responsible for sampling decisions. 
In tracing systems, different mechanisms exist to decide if a trace should be collected. 
However, since all tracing logic must be executed in a function, the mechanisms are limited. 
Specifically, only event-based and probability-based sampling are possible, as all other approaches require additional communication with external systems.

More detailed documentation of the tracing module is available in the public repository\footnote{\url{https://github.com/flamestro/DistributedTracingServerlessExample}}.

\subsection{Platform-supported Tracing}
As described in previous sections, we implemented platform-supported tracing in OpenWhisk, as it is one of the only existing open-source FaaS-platform to evaluate these types of augmentations. 
To ensure reproduction of our contribution, we based our augmentation on commit \texttt{6928a1d} of the OpenWhisk project\footnote{\url{http://openwhisk.apache.org/}}.

Coincidentally, the first platform-supported tracing approach was already proposed in the OpenWhisk GitHub on Jun 22, 2017. Thus, we started our implementation based on the existing Pull-Request \#2413\footnote{\url{https://github.com/apache/openwhisk/pull/2413}}. However, we needed to extend the existing Pull-Request to allow for full platform-supported tracing. Specifically, the solution does not pass the  \texttt{trace context} to the action. Thus, it can observe platform events but not platform and application events.
Additionally, the existing Pull-Request is not able to capture inter-dependencies in function calls. 
Our platform augmentation enables developers to see function executions and all OpenWhisk components, e.g., Invoker, Controller for a specific trace. Thus, both platform and application (function) faults are observable in one consolidated view.

The implementation of platform-supported tracing includes the addition of Zipkin to the OpenWhisk components. Thus, a provider of OpenWhisk must now also manage Zipkin and allow developers to access Zipkin. 
We mainly used the OpenTracing Scala library that provides all the functionalities that the platform needs to communicate with a tracing backend and create the trace.
However, as mentioned above, we need to enable the application developer to attach additional tracing information to the activation. 
We augmented the OpenWhisk controller to manage the \texttt{trace context} in the controller, as only the controller knows if a function invocation is related to a specific event.
Thus, an invoker can be aware of the current context and inject it into activation through an environment variable. Besides changes to the Invoker and Controller, we also needed to change the runtime implementation to capture and interpret the \texttt{trace context} for each activation. For more details, see the GitHub for the provider-side augmentations\footnote{\url{https://github.com/flamestro/openwhisk}}. Lastly, we needed function-side support to interpret and use the injected \texttt{trace context}. We can use the already discussed developer-driven module, but a developer can also choose to implement that themselves.

\subsection{Summary}
Both approaches enable a developer to observe application-side events and thus capture application faults. However, the developer must deploy and manage the tracing backend and instrument clients to trigger trace sampling in the developer-driven approach. On the other hand, the platform-supported tracing offers a tracing backend as an additional feature to developers. However, this also implies overhead for platform providers in management and security. This also means that developers require additional network resources to perform application-side tracings as it is the platform's responsibility to collect and aggregate this information. In the next section, we evaluate both approaches in terms of performance and cost for both providers and developers. We also discuss the level of detail a developer can observe using each method.

%% file: sections/05_eval.tex
This section evaluates both implementations concerning fault observability and associated trade-offs in terms of execution latency and resource utilization.
First, we use the model demonstrated in Table~\ref{tab:fault-observability-tracing} to characterize the trace-observable evidence generated by the two approaches and highlight differences. Then, we describe a preliminary experiment to measure the performance and cost trade-offs of both tracing approaches in OpenWhisk.

\subsection{Evaluation of Trace-observable Evidence}
For the evaluation of observability, we reuse the four faults presented in section~\ref{sec:appliedmodel}. 
This time, we compare AWS X-Ray against the two tracing approaches we developed for OpenWhisk. Table \ref{tab:fault-observability-tracing} shows our results. 

\begin{table*}[tp]
\caption{
Fault observability of distributed tracing approaches; underlining indicates improvements over response/log observability, brackets indicate partial improvements}
\input{data/faultobservability_tracing_v02}
\label{tab:fault-observability-tracing}
\end{table*}

Unsurprisingly, all faults generate trace-observable evidence in case of a fault. Other improvements over log-observable evidence are underlined in the table. In the following, we highlight the most important differences: 

Starting with \texttt{F1}, we see that distributed tracing with all approaches makes this type of fault visible and remains consistent with log-observability.
For \texttt{F2}, we determined that the evidence generated by AWS X-Ray is able to show the lagging request and therefore provide unambiguous evidence. AWS provides SDKs that automatically instrument downstream calls by patching some of the most used libraries. This includes the libraries needed for API calls. Through automatic instrumentation, developers can trace faults even in the most negligent code that lacks error handling. 
Both OpenWhisk approaches can also provide unambiguous evidence for this fault as long as the code is instrumented in the API call.

Fault \texttt{F4} shows that distributed tracing can clear up evidence inconsistencies for async functions in all three approaches. When a fault occurs in the asynchronous function, the span will be tagged as such. A look at the entire trace of the execution paths reveals this fault in both provider-supported tracing and developer-driven tracing because the fault happened inside the function logic. 

Lastly, we look at fault \texttt{F3}. In order to provide unambiguous evidence here, traces have to show not only the dependency between the two functions but also explain the variance in latency that sometimes leads to a timeout. AWS X-Ray and the platform-supported approach of OpenWhisk distinguish between Initialization and Invocation of functions in their traces. Developer-driven tracing is implemented as part of the function code, so it can only trace the execution inside functions and not the execution steps between two functions and therefore does not provide unambiguous and consistent evidence. %

In summary, we can see that developer-driven tracing is already able to improve consistency (\texttt{F4}) and, in some cases, reduce ambiguity (\texttt{F2}). Platform-supported tracing reduces ambiguity by giving the developer more insight into faults caused by the platform (\texttt{F3}) and is comparable to Amazon's XRay.
Thus, the only remaining question we need to ask is: What performance and cost trade-offs does platform-supported tracing make to achieve this improvement is observability?

\subsection{Experiment Design}
In the following, we present a preliminary experiment design that evaluates the impact of both implementations in terms of execution-time overhead and platform overhead. 
The experiment aims to give platform designers and application engineers a first insight into the cost of integrating distributed tracing with either approach.
As a use-case, we selected a batch-import process presented in Figure~\ref{fig:app}.

\paragraph{System Under Test}
To enable the experiment's reproduction, we describe how the system and the configuration around the experiment are set up in the following.
We evaluate our implementation on OpenWhisk version 4.2.2 using the default Python 3 runtime provided by OpenWhisk. Specifically, our experiment is based on the commit \texttt{6928a1d} of OpenWhisk. 
The experiment was conducted on a four-node Kubernetes cluster hosted at TU Berlin.
For each function in our use-case example, we created a single zip file containing the function code and all its dependencies.
We deployed our use-case example (figure~\ref{fig:app} in three variants, one without any tracing enabled, one with developer-driven tracing and one with platform-supported tracing. 
The memory limit of all functions is set to 128 MB with a timeout of 300 seconds. The application responses to a HTTP request that links to a prepared CSV file (application workload).

\paragraph{Fault Injection}
In our experiment, we dynamically inject faults based on pre-defined occurrence chances.
\begin{itemize}
	\item trigger an uncaught exception
	\item causing timeouts
	\item import an invalid dependency causing a runtime error.
	\item causing a system error, by killing a function container
\end{itemize}
As part of the experiment, we created a small python chaos testing module for calling functions with the ability to insert each of the runtime-side errors. We performed the system error injection manually.

\paragraph{Workloads and Defect Mapping}
We use a CSV file containing 150 records for our experiment, each with one or more product image references.
The used CSV file will trigger at least 303 invocations, including at least 150 invocations of the \texttt{fetchProductImages}-function.
For each application variant, we make 100 import requests.

\paragraph{Measurement Points}
In order to observe both the accuracy, execution time and platform costs, we measure each function's execution time, the overall import time, the CPU, memory and network utilization of all four Kubernetes nodes. Further, we use methods presented in \cite{kuhlenkamp-2020-sac-elasticity,kuhlenkamp-2019-ucc-opstasks} to detect cold-starts and other function-side metrics. 
For each request, we also collected the injected fault type.

Based on these measurements, we can extract the accuracy of the traces by manually checking if the injected fault is observable. We used these experiments to complete table~\ref{tab:fault-observability-tracing}.

\subsection{Experiment Results}
The costs of developer-driven and platform-supported tracing are split into client-side costs and provider-side costs. 
The client-side costs are reflected by the execution time of all functions in the serverless application.
The provider-side costs, on the other hand, are reflected by resource utilization.

Thus, we evaluate the cost of implementing and using the tracing system from a client perspective, as each tracing request will create an additional network and compute overhead per function. 
For the provider-side cost, we measure the added CPU, memory usage resulting from processing and producing traces. For the provider-supported tracing approach, we are particularly interested in how much resources are lost on the worker nodes, as this impacts how many functions we can ultimately run on a worker node and thus monetize. 

\paragraph*{Client-side Cost}
We did the client-side cost evaluation based on the execution times of the \texttt{fetchProductImages} function because the \texttt{fetchProductImages} function creates a high amount of spans and is the longest-running function in our application, see figure~\ref{fig:app}. We assume that the client-side impact of tracing instrumentation is the same of all functions. We present the client-side cost in comparison to the baseline in Table~\ref{tab:client-side-cost}.

\begin{table}[b]
\centering
\caption{Aggregated execution time of fetchProductImages in comparison to the baseline (no instrumentation and tracing)}
\label{tab:client-side-cost}

\input{data/client_costs}
\end{table}

Based on this evaluation, we can observe that the function execution time for developer-driven tracing does not increase on average. The maximum execution time increases by more than a minute. 
This is mainly caused by the added network interaction with the tracing system and by the integration of the necessary dependencies. Both these issues could be addressed by implementing serverless specific tracing libraries, an optimization outside of the scope of a research prototype.
For platform-supported tracing, we see an increase in mean execution time of 2 seconds and a maximum execution time of 82 seconds over the baseline. 
However, we should note that the used sampling rate of 100\% which leads to this noticeable impact, is not typical for a production tracing system. 
The stress on platform components would be significantly less if the sampling would adapt to application behaviour instead. Besides, the deployment of the tracing infrastructure could be improved, e.g., each invoker-node could use a Zipkin proxy to offload connection overhead. Thus, our experiment shows the most extreme case.

\paragraph*{Provider-Side Cost}

We did the provider-side cost evaluation based on the platform's resource utilization during the experiments. Specifically, we observed the CPU and memory utilization. We present these costs in terms of available execution units per second. We define an \textit{execution unit} as the smallest possible function container in our deployment, which is a 128 MB runtime using one CPU unit. Our four-node cluster has a maximum of 96 free CPU units.

Based on our observation of 100 invocations of the bulk import application, we observed no impact on the developer-driven implementations. The CPU utilization showed a slight increase, implying that we might not be able to pack all 96 CPU units without risking noisy neighbour effects. This is likely due to the added network usage in comparison to the baseline. 
For the platform-supported side implementation, we observed a different story. Here we see that between 128 and 256 MB additional memory is used. This overhead is likely due to the instrumentation in the invoker and controller. Thus, a provider implementing platform-supported tracing might lose between 1 and 2 runtimes when offering this feature. Notably, the implementation could likely be improved and optimized further to reduce this impact.

In summary, we can see that both implementations impact performance overhead marginally while improving trace observability by reducing fault ambiguity and fault inconsistencies. 
We further see that developer-driven tracing can improve observability when platforms do not provide tracing support with marginal performance impacts and thus should be considered for all serverless applications.

%% file: data/faultobservability_tracing_v02.tex
\centering

\begin{tabular}{|c|ccc|ccc|ccc|}
\hline
\multirow{3}{*}{Platform / Fault  } & \multicolumn{3}{c|}{AWS} & \multicolumn{6}{c|}{OWhisk} \\ \cline{2-10} 
 & \multicolumn{3}{c|}{XRay} & \multicolumn{3}{c|}{Developer-driven Tracing} & \multicolumn{3}{c|}{Platform-supported Tracing} \\
 & visible & unambiguous & consistent & visible & unambiguous & consistent & visible & unambiguous & consistent \\ \hhline{|==========|}
F1 & true & true & true & true & true & true & true & true & true \\
F2 & true & true & \uline{true} & true & true & \uline{(true)} & true & true & \uline{true} \\ 
F3 & true & \uline{true} & \uline{true} & true & \textbf{false} & \textbf{false} & true & \uline{true} & \uline{true} \\
F4 & true & \uline{true} & \uline{true} & \uline{(true)} & \uline{(true)} & true & true & \uline{true} & \uline{true} \\ \hline
\end{tabular}

%% file: data/client_costs.tex
\begin{tabular}{|c|r|r|r|}
\hline

 & \multicolumn{3}{c|}{$\Delta baseline$} \\ \cline{2-4} 

\multirow{-2}{*}{\parbox{2.5cm}{Function Execution Overhead}} & \multicolumn{1}{c|}{mean [s]} & \multicolumn{1}{c|}{median [s]} &
\multicolumn{1}{c|}{max [s]} \\
\hhline{|====|}
Developer-driven & 0 & 3 & 77 \\ \hline
Platform-supported & 2 & 4 & 82 \\ \hline
\end{tabular}%

%% file: sections/06_relatedwork.tex
The problem of troubleshooting or debugging serverless applications is well documented in the literature.  Serverless environments are not easily replicated, so more often than not, debugging happens in production \cite{Spillner_ServerlessFutureChallenges_2021}. When errors occur in complex service compositions, they can be more complicated to identify than in traditional cloud models~\cite{kuhlenkamp-2020-ic2e-ifs_and_buts}.

To address this issue, Manner et al.~\cite{2019-Manner-SICS-FaasDebugging} suggest a concept that relies on alerts and appropriate log messages to provide a semi-automatic troubleshooting process. This approach combines monitoring and logging but does not investigate the use of distributed tracing. Indeed, the use of distributed tracing in serverless applications has barely been investigated. 

An exception here is the work of Lin et al.~\cite{Lin_TracingLambdaDependencies_2018}. Similar to our work, the researchers compare different approaches for distributed tracing. However, the comparison only considers overhead, sampling and instrumentation effort and does not characterize or compare the observability of the different approaches.
Lenarduzzi and Panichella~\cite{Lenarduzzi_ServerlessTestingTools_2021} interview different developers to explore serverless debugging tools. They endorse the use of distributed tracing, but do not compare tools or provide specific recommendations. 
Kuhlenkamp et al.~\cite{kuhlenkamp-2018-icsoc-costradamus} implement a developer-driven tracing approach to trace the costs of serverless applications but do not address the problem of faults with their implementation. 
Besides these contributions, little has been researched about improving the observability of serverless applications in particular, as also noted by~\cite{2019-Manner-SICS-FaasDebugging}.

In the more general field of microservice architectures however,  observability is increasingly becoming a relevant research topic. 
For the use-case of fault analysis in particular, the researchers in~\cite{Zhou_FaultAnalysisTracingMSA_2018} contribute an industrial survey and experiment. They analyse the benefits of distributed tracing by measuring the time for the different debugging practices. Most of the faults investigated relate to developer bugs in particular and not platform influences that are typical in serverless applications.
More generally, Niedermaier et al.~\cite{Niedermaier_ObservabilityInterviewStudy_2019} also survey challenges and good practices in the field of observability and monitoring of distributed systems.

%% file: sections/07_conclusion.tex
Our experiments show that the integration of tracing in serverless applications increases observability, especially within function interaction. We further see that platform providers can easily improve the observability for transmission and system errors by adding platform-supported tracing, which gives the developer tracing access to the entire execution process. 

Further, we showed that the overhead of integrating tracing inside an application is negligible in terms of execution overhead. Additionally, the overhead for offering platform-supported tracing even at the highest granularity only causes the loss of a portion of the resources needed for a function runtime. Further optimization in the tracing implementation in the platform might reduce this even further. We did not evaluate the extra cost of providing the tracing system itself. However, any cloud provider already offers some form of distributed tracing as part of their infrastructure. Thus the existing cost model for these systems can be used to offset the cost of offering tracing.
Lastly, we presented a serverless observability model, highlighting how faults' different visibilities can cause inconsistencies and ambiguities for pinpointing faults. 

In the future, we plan to extend this model to other distributed application models and investigate system designs and tools to address the inconsistencies and ambiguities in modern cloud-native applications.

%% file: main.bbl
\begin{thebibliography}{10}
\bibitem{Spillner_ServerlessFutureChallenges_2021}
D.~{Taibi}, J.~{Spillner}, and K.~{Wawruch}, ``Serverless computing-where are
  we now, and where are we heading?'' \emph{IEEE Software}, vol.~38, no.~1, pp.
  25--31, 2021.
\end{thebibliography}
